\documentclass[%
 aip,
 amsmath,amssymb,
 reprint,%
]{revtex4-1}

\usepackage{graphicx}% Include figure files
\usepackage{dcolumn}% Align table columns on decimal point
\usepackage{bm}% bold math
\usepackage[utf8]{inputenc}
\usepackage[T1]{fontenc}
\usepackage{mathptmx}

\usepackage{natbib}
\usepackage{amsfonts}
\usepackage{mathrsfs}
\usepackage[english]{babel}
\usepackage{epsfig}
\usepackage{epstopdf}
\usepackage{babelbib}
\usepackage{hyperref}
\usepackage{color}
\usepackage{tcolorbox}
\usepackage{tabularx}
\usepackage{array}
\usepackage{colortbl}
\tcbuselibrary{skins}
\usepackage{comment}
\usepackage{array}
\usepackage{mathtools}
\usepackage[utf8]{inputenc}
\usepackage{afterpage}

\makeatletter
\newcommand*{\textoverline}[1]{$\overline{\hbox{#1}}\m@th$}

\makeatother

\begin{document}

\preprint{AIP/123-QED}

\title{Global order parameters for particle distributions on the sphere}
% Force line breaks with \\

\author{A. Bo\v zi\v c}
\affiliation{Department of Theoretical Physics, Jo\v{z}ef Stefan Institute, SI-1000 Ljubljana, Slovenia}
\email{anze.bozic@ijs.si}
\author{S. Franzini}
\affiliation{Scuola Internazionale Superiore di Studi Avanzati, I-34136 Trieste, Italy}%
\author{S. \v Copar}
\affiliation{Faculty of Mathematics and Physics, University of Ljubljana, SI-1000 Ljubljana, Slovenia}

\date{\today}% It is always \today, today,

\begin{abstract}
Topology and geometry of a sphere create constraints for particles that lie on its surface which they otherwise do not experience in Euclidean space. Notably, the number of particles and the size of the system can be varied separately, requiring a careful treatment of systems with one or several characteristic length scales. All this can make it difficult to precisely determine whether a particular system is in a disordered, fluid-like, or crystal-like state. Here, we show how order transitions in systems of particles interacting on the surface of a sphere can be detected by changes in two hyperuniformity parameters, derived from spherical structure factor and cap number variance. We demonstrate their use on two different systems---solutions of the thermal Thomson problem and particles interacting via an ultra-soft potential of the generalized exponential model of order $4$---each with a distinct parameter regulating their degree of ordering. The hyperuniformity parameters are not only able to detect the order transitions in both systems, but also point out the clear differences in the ordered distributions in each due to the nature of the interaction leading to them. Our study shows that hyperuniformity analysis of particle distributions on the sphere provides a powerful insight into fluid- and crystal-like order on the sphere.
\end{abstract}

\maketitle

\section{\label{sec:intro}Introduction}

Geometrically frustrated assemblies are ubiquitous in biological, soft, and condensed matter~\cite{Grason2016}, yet even the influence of spherical geometry---perhaps the simplest closed, curved surface---on crystallization and ordering of particles remains poorly understood~\cite{Manoharan2015,Giarritta1992,Guerra2018,Chen2020}. It is important to understand and determine the degree of (dis)order in spherical structures, as it can lead to different optic~\cite{Vogel2015,Jacucci2020}, elastic~\cite{Yong2013,Brojan2015}, and dynamic~\cite{Yao2019} properties. Crystal-like order and defects have been studied in viruses~\cite{Wang2018a,Zandi2020}, metazoan epithelia~\cite{Roshal2020}, and colloidal capsules~\cite{Fantoni2012,Thompson2015,Bollhorst2017}, where different ways of construction have been shown to lead to different degrees of order~\cite{Wang2018}. At the same time, the order of the underlying spherical lattice can also significantly influence the orientations of anisotropically interacting particles positioned on it~\cite{Gnidovec2020,Copar2020,Gnidovec2021}. Order parameters describing local positional relationships between particles are also an important tool in determining structural features of liquids and glasses~\cite{Tanaka2019}. In such disordered systems, particle arrangement tends to be much better studied on a local scale (shells of nearest neighbours) than on larger scales~\cite{Royall2015,Hallett2018}. However, order parameters which can detect long-range structure and changes in it are useful in studies of liquids exhibiting a shell-like structure with changing order~\cite{Markovich1996,Sharma2006,Royall2017,Zhang2020} as well as in changes in shape in cooperatively rearranging regions in colloidal liquids on a sphere~\cite{Singh2020}.

Despite its importance, it can be difficult to characterize the degree of order in a distribution of particles on the sphere, particularly given the numerous defects and topological scars present even in the most ordered structures~\cite{Bowick2009,CCD2}. Often, local bond order parameters~\cite{Steinhardt1983} are used to detect the presence and onset of order in particle distributions on the sphere~\cite{Li2013,Guerra2018,Vest2018,Mosayebi2017}, but they tend to be based on the expected order and the prevalent 6-fold character of a locally-ordered crystal-like particle distributions. Other (``order-agnostic''~\cite{Royall2015}) measures such as mesh ratio and energy are used to distinguish between different types of spherical structures~\cite{Hardin2016}, but there are important exceptions where neither these nor local bond order parameters can provide a good answer~\cite{Franzini2018,Mickel2013}. Recently, however, some progress has been made by extending the notion of hyperuniformity, thoroughly explored in Euclidean space~\cite{Torquato2003,Torquato2018}, to the sphere and other curved surfaces~\cite{ALB2019,Lomba2020,Meyra2019,Brauchart2018,Brauchart2020,Stepanyuk2020}, which introduces a more global view of the order on the sphere.
    
In our previous work~\cite{ALB2019} it was shown that by extending the notion of hyperuniformity to spherical geometry, it is possible to derive two parameters which together indicate the degree of order in scale-free distributions of particles on the sphere. Here, this notion is generalized to systems with one or more length scales, which are more difficult to tackle, as both the number of particles and the size of the system have to be treated as independent parameters~\cite{Hill1994,Post1986}. Nonetheless, we demonstrate that the hyperuniformity parameters can be used to detect the degree of order even in those systems where usual approaches fail. Hyperuniformity on a sphere could thus provide a good framework for a consistent definition of fluid- and crystal-like order on the sphere.

\section{Methods}

\subsection{Spherical structure factor and cap number variance}

We describe an arbitrary distribution of $N$ particles on the surface of a sphere with radius $R$ with a surface density distribution
\begin{equation}
\rho(\Omega)=\frac{1}{R^2}\sum_{k=1}^N\delta(\Omega-\Omega_k)=\frac{1}{R^2}\sum_{\ell,m}\rho_{\ell m}Y_{\ell m}(\Omega),
\end{equation}
where $\Omega_k$ are the positions of the particles in spherical coordinates $(\vartheta,\varphi)$. The coefficients $\rho_{\ell m}$, used to expand the distribution in terms of spherical harmonics $Y_{\ell m}(\Omega)$, further define the spherical structure factor~\cite{ALB2019},
\begin{equation}
\label{eq:sf}
S_N(\ell)=\frac{1}{N}\frac{4\pi}{2\ell+1}\sum_m\left|\rho_{\ell m}\right|^2=\frac{1}{N}\sum_{i,j=1}^NP_\ell(\cos\gamma_{ij}).
\end{equation}
Here, $P_n(x)$ are the Legendre polynomials and $\gamma_{ij}$ is the spherical distance between particles $i$ and $j$. Spherical structure factor is tightly related to the pair correlation function~\cite{Franzini2018,Viveros2008,Vest2018} and should reflect the interaction potential of the system. The spherical structure factor can be connected to another measure, the cap number variance $\sigma_N^2(\theta)$, which gives the variance of the number of particles contained in a spherical cap with an opening angle $\theta$~\cite{Brauchart2018,ALB2019}:
\begin{equation}
\label{eq:sig}
\sigma_N^2(\theta)=\frac{N}{4}\sum_{\ell=1}^\infty S_N(\ell)\frac{\left[P_{\ell+1}(\cos\theta)-P_{\ell-1}(\cos\theta)\right]^2}{2\ell+1}.
\end{equation}
In practice, $\sigma_N^2(\theta)$ is obtained by covering the sphere with a series of randomly-positioned spherical caps with an opening angle $\theta$, determining the number of particles in each, and calculating their variance.

\subsection{Hyperuniformity on the sphere}

It has been shown previously~\cite{ALB2019} for scale-free particle distributions on the sphere that the form of the cap number variance in Eq.~\eqref{eq:sig} can be approximated by
\begin{equation}
\label{eq:cnv}
\sigma_N^2(\theta)=A_N\frac{N}{4}\sin^2\theta+B_N\frac{\sqrt{N}}{4\sqrt{3}}\sin\theta,
\end{equation}
with an additional (small) residual, relevant only in the case of ordered distributions. The form of cap number variance in Eq.~\eqref{eq:cnv} can be considered a spherical analogue of the asymptotic form of the number variance in Euclidean space, used to determine the degree of hyperuniformity in such systems~\cite{ALB2019,Torquato2018}. Furthermore, the two parameters in Eq.~\eqref{eq:cnv}, $A_N$ and $B_N$, turn out to be particularly good measures of order in scale-free particle distributions on the sphere. For a completely random distribution, one can show that $A_N=1$ and $B_N=0$; this corresponds to a uniform structure factor where $S_N(\ell)=1$ $\forall\ell$. With a gradual onset of order in a system of particles, first a low-$\ell$ gap starts to appear in the structure factor, and simultaneously $A_N$ starts to diminish while $B_N$ increases. In the limit of $A_N\to0$, equivalent to the onset of hyperuniformity in Euclidean space, a particle distribution on the sphere becomes ordered with a series of pronounced peaks in its structure factor and can be, in principle, characterized by its value of $B_N$---which depends not only on the type of distribution but also on any symmetries present in it (for details, see Ref.~\onlinecite{ALB2019}).

\subsection{Distributions of particles on the sphere}

In this work, we generalize these results to particle distributions with one or more internal length scales. To do this, we study two completely different systems: {\em (i)} solutions of the thermal Thomson problem, where temperature introduces a length scale into an otherwise scale-free system; and {\em (ii)} particles interacting via an ultra-soft potential of the generalized exponential model of order $4$ (GEM-4), where the system exhibits an ordered phase of cluster crystals depending on both the number of particles and the size of the system.

In the first case, {\em (i)}, particles interact via long-range electrostatic potential, just as in the classical Thomson problem, but we additionally introduce a temperature $T$ into the system. The $N$ particles all carry identical charge, and the electrostatic energy of the system is determined by pairwise summation, $V=E_0\sum_{i> j}|{\bm r}_i-{\bm r}_j|^{-1}$, where $E_0=e^2/4\pi\varepsilon_0R$ sets the interaction scale and the distance is measured in the Euclidean sense. Here, $\varepsilon_0$ is the vacuum permittivity and $R$ the sphere radius. We work with dimensionless units and we use a reduced temperature for the system, $T=k_BT^*/E_0$, where $T^*$ is the real temperature and $k_B$ is the Boltzmann constant.

To obtain ensembles at different temperatures, we start at a very high temperature, $T=10^3$, and gradually lower it. This is achieved by virtue of Monte Carlo simulations~\cite{Frenkel,Xiang1997}, where at each temperature step we perform a series of random displacements of individual particles drawn from a spherical Gaussian (von Mises-Fisher) distribution centred around a particle~\cite{ALB2018b}. For the width parameter of the distribution, we choose $\lambda=\sqrt{N}/T$, which ensures a good acceptance rate also at low $T$ and large $N$. We have also tested a few other choices of $\lambda$, which turn out to work similarly well. After a burn-in phase, configurations are sampled every $4N$ moves until $250$ different configurations are obtained, which are then used to obtain ensemble-averaged spherical structure factor and cap number variance. At the lowest temperatures studied, $T\gtrsim10^{-5}$, the system converges to the known minima of the Thomson problem. Nonetheless, due to the nature of the procedure, the system can for a given $N$ get trapped in a local minimum, whose energy remains very close to the known energy minima of the Thomson problem~\cite{CCD2} (with a relative error of $\lesssim10^{-5}$).

In the second case, {\em (ii)}, we study particles interacting via a generalized exponential model of order 4 (GEM-4 potential~\cite{Franzini2018}): a bounded, purely repulsive soft pair potential of the form $w(r)=\varepsilon\exp(-(r/\delta)^4)$. Here, $\varepsilon$ and $\delta$ determine the energy and length scales of the model. We use the former to again rescale the temperature $T$ of the system, $T=k_BT^*/\varepsilon$, while the latter introduces a length scale to the system, $\delta/R$, which controls its phase behaviour. The distance between the particles $r$ is measured along the surface of the sphere. While the GEM-4 system is also simulated at a finite $T$, this is the least interesting variable in the system; we will thus study the system at  $T=1$, unless specified otherwise, and explore its behaviour with respect to both $N$ and $\delta/R$. At each point in the phase space,  $50$ configurations are sampled to generate ensemble-averaged spherical structure factor and cap number variance. Further simulation details can be found in Ref.~\onlinecite{Franzini2018}.

%\section{Results}

\section{Thermal Thomson problem}

Solutions of the Thomson problem are distributions of particles minimizing their electrostatic interaction~\cite{Wales2006}. Known minimum energy distributions are often characterized by a high symmetry and a locally triangular mesh where each particle has 6 neighbours, with the exception of 12 5-fold defects owing to the topology of the sphere; at high $N$, pairs of defects in the form of topological scars are also common~\cite{Bowick2009}. When temperature (measured relative to the interaction energy) is introduced into the system, the order disappears and different kinds of local defects are ubiquitous. Temperature is also the only length scale in the system---note that changing the radius of the sphere is equivalent to changing the scale of interaction energy and thus the scale of the reduced temperature. As it is lowered, the solutions of the thermal Thomson problem converge towards the known minima of the Thomson problem.

\begin{figure}[!b]
\centering
\includegraphics[width=\columnwidth]{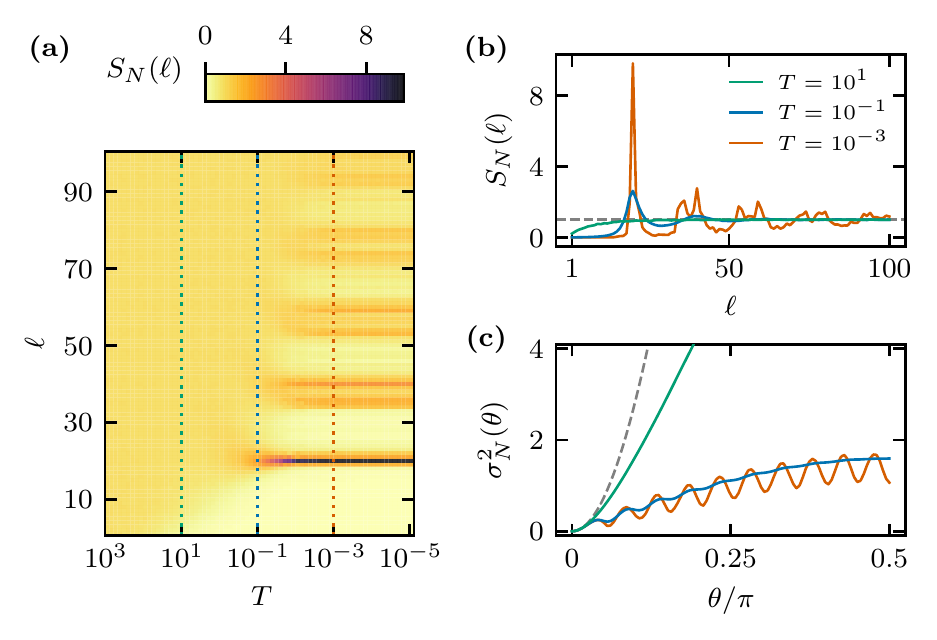}
\caption{{\bf (a)} Ensemble-averaged spherical structure factor for distributions of $N=120$ particles in the $T$--$\ell$ plane. Note that the $x$ scale is inverted to reflect the onset of order as temperature is decreased. Ensemble-averaged spherical structure factor {\bf (b)} and cap number variance {\bf (c)} of distributions at three different temperatures, marked by dotted lines in panel (a). The legend of panel (b) also applies to panel (c). Dashed gray lines show the expected behaviour for a completely random distribution: $S_N(\ell)=1$ $\forall\ell$ [panel (a)] and $\sigma_N^2(\theta)=N\sin^2\theta/4$ [panel (b)].\label{fig:1}}
\end{figure}

\subsection{Structure factor and number variance}

The temperature-dependent order transition in a thermal Thomson system can be easily observed when we take a look at the (ensemble-averaged) spherical structure factor [Eq.~\eqref{eq:sf}]. Figure~\ref{fig:1}a shows $S_N(\ell)$ in the $T$--$\ell$ plane for distributions of $N=120$ particles. At high $T$, the structure factor is essentially indistinguishable from that of a random distribution, $S_N(\ell)=1$ $\forall\ell$. As $T$ is lowered, $S_N(\ell)$ becomes progressively more defined: first, a gap appears at low $\ell$, growing with decreasing temperature, and as it approaches $\ell_0\approx\pi\sqrt{N}/\sqrt{3}$~\cite{ALB2019}, the first peak of the structure factor appears at $\ell_0$ (Fig.~\ref{fig:1}b). Its position does not change as $T$ is lowered further; on the other hand, higher-$\ell$ peaks do not initially appear at the exact positions of the crystal-like (minimum energy) state, but shift slightly with decreasing $T$. At the very lowest $T$, the form of $S_N(\ell)$ is completely defined and approaches the form of the known minimum solutions of the Thomson problem~\cite{ALB2019,CCD2}. In some cases, discrepancies remain: these structures, while ordered and crystal-like, are trapped in local minima.

Spherical structure factor is directly related to cap number variance [Eq.~\eqref{eq:cnv}], the variance in the number of particles contained in spherical caps with opening angle $\theta$. As the temperature of the system is lowered and $S_N(\ell)$ becomes more defined, the angular dependence of $\sigma_N^2(\theta)$ goes from $\propto\sin^2\theta$, characteristic of a random distribution, to $\propto\sin\theta$, typical of crystal-like distributions~\cite{ALB2019}. Furthermore, when the order in the distribution becomes crystal-like, $\sigma_N^2(\theta)$ also starts to exhibit a modulation on top of its general $\theta$-dependence, whose form is related to $\ell_0$ (and thus to $N$) and is another consequence of ordering~\cite{ALB2019}.

\subsection{Hyperuniformity parameters}

Changes in spherical structure factor and cap number variance can be summarized by fitting $\sigma_N^2(\theta)$ to the form  given by Eq.~\eqref{eq:cnv}, which yields two hyperuniformity parameters $A_N$ and $B_N$. As already mentioned, it has been shown previously~\cite{ALB2019} that for a completely random distribution, $A_N=1$ and $B_N=0$, while on the other hand, ordered distributions (such as minima of the Thomson and Tammes problems) have $A_N=0$ and $B_N\lesssim1$. This, of course, holds in the average sense, particularly for randomly-generated distributions.

\begin{figure}[!tb]
\centering
\includegraphics[width=\columnwidth]{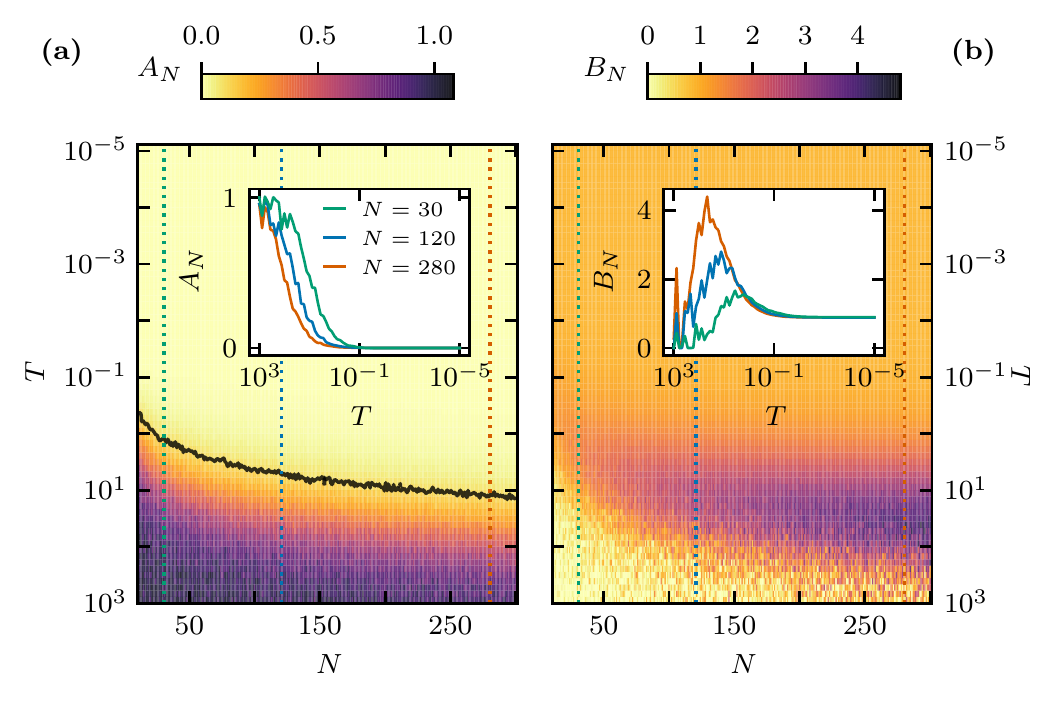}
\caption{Hyperuniformity parameters $A_N$ {\bf (a)} and $B_N$ {\bf (b)} of thermal Thomson distributions in the $N$-$T$ plane, obtained as fits of Eq.~\eqref{eq:cnv} to ensemble averages of cap number variance $\sigma_N^2(\theta)$. The $y$ scale is inverted in both cases to reflect the onset of order as temperature is decreased. Insets show the temperature dependence of the two parameters for three different values of $N$, marked in the main plots with dotted lines. The legend of the inset in panel (a) also applies to the inset in panel (b). Black contour line in panel (a) shows the critical temperature $T_c$ where $A_N\leqslant0.1$, which is proportional to $T_c\propto N$.\label{fig:2}}
\end{figure}

Figure~\ref{fig:2} shows the hyperuniformity parameters $A_N$ and $B_N$ of the thermal Thomson distributions in the $N$-$T$ plane. The fits of Eq.~\eqref{eq:cnv} are performed on ensemble-averaged curves $\sigma_N^2(\theta)$ for each $N$ and $T$, as fits to individual ensemble samples do not yield reliable results due to the large degree of randomness present in the system. We can observe several things: at high $T$ when the system is disordered, $A_N\lesssim1$ and $B_N\approx0$, close to the values pertaining to random distributions (albeit not completely, as even at highest $T$ the system is not completely random due to the interactions involved; see Fig.~\ref{fig:1}). Furthermore, higher $N$ have lower values of $A_N$ at high $T$, which is understandable since the energy of the system also increases with $N$. As $T$ is lowered, $A_N\to0$ and we can talk about the onset of crystal-like order in the distributions~\cite{ALB2019}. The transition is gradual, and the critical temperature shows a slight dependence on $N$---for higher $N$, the transition occurs at higher $T$. The approximate range of temperatures where the transition occurs is $T\sim1$ to $0.1$ (insets of Fig.~\ref{fig:2}).

Since the order transition is continuous with temperature, we cannot speak of a clear critical temperature of the onset of order. However, a reasonable threshold $c$ can be chosen so that $A_N\leqslant c$, since we know that $B_N$ already has a peak when $A_N$ is of the order of magnitude of a few tenths (Fig.~\ref{fig:2}). Any threshold choice above $c\gtrsim10^{-3}$ shows almost exact proportionality of the critical temperature to the number of particles, $T_c\propto N$ (Fig.~\ref{fig:2}a shows the example of $c=0.1$). On the other hand, setting the threshold to even lower values is too affected by noise to be suitable for defining a transition temperature. The increasing trend reflects stronger bonding and thus higher energy cost of displacement when particles are packed closer together. In a physical system, the energy scale also includes the size of the sphere and the particle charges which were set in our case.

Parameter $B_N$ is dominated by noise at high $T$, since $A_N$ is the dominant parameter there. As $T$ is lowered, $B_N$ typically crosses a ``barrier'' in the temperature range where $A_N$ first starts to decrease, the height of which increases with $N$. This increase in $B_N$ as $A_N$ is lowered could thus indicate some particular property of the interactions in the system. At low $T$ where $A_N\to0$, values of $B_N$ start to converge to very similar values regardless of $N$, $B_N\sim0.9$, which is characteristic of ordered distributions on the sphere in general and minimum solutions of the Thomson problem in particular~\cite{ALB2019}. The vanishing of parameter $A_N$ thus clearly signals a transition from a disordered to an ordered distribution. The parameter $B_N$, on the other hand, becomes relevant only when $A_N$ vanishes---then, $B_N$ carries some information about the nature of the order.

\section{GEM-4 potential}

The same analysis that has been done for the thermal Thomson problem can be applied to particle distributions resulting from the GEM-4 interaction potential. At a given $T$ (note that the temperature related to the GEM-4 potential has a different scale than the one pertaining to the thermal Thomson problem), the system of GEM-4 particles is known to undergo an ordering transition from a homogeneous fluid to a cluster crystal phase, depending on both the number of particles $N$ and the (scaled) radius of the sphere $\delta/R$~\cite{Franzini2018}. At high density, particles aggregate into clusters at sites which are distributed on the sphere in a highly ordered manner (Fig.~\ref{fig:3}a). However, the internal structure of such clusters remains disordered, as particles randomly move inside the potential well. The number of clusters is a function of $\delta/R$ but not $N$---an increase in the number of particles at a fixed sphere size will only lead to each cluster having more particles.

\subsection{Structure factor and number variance}

\begin{figure}[!b]
\centering
\includegraphics[width=\columnwidth]{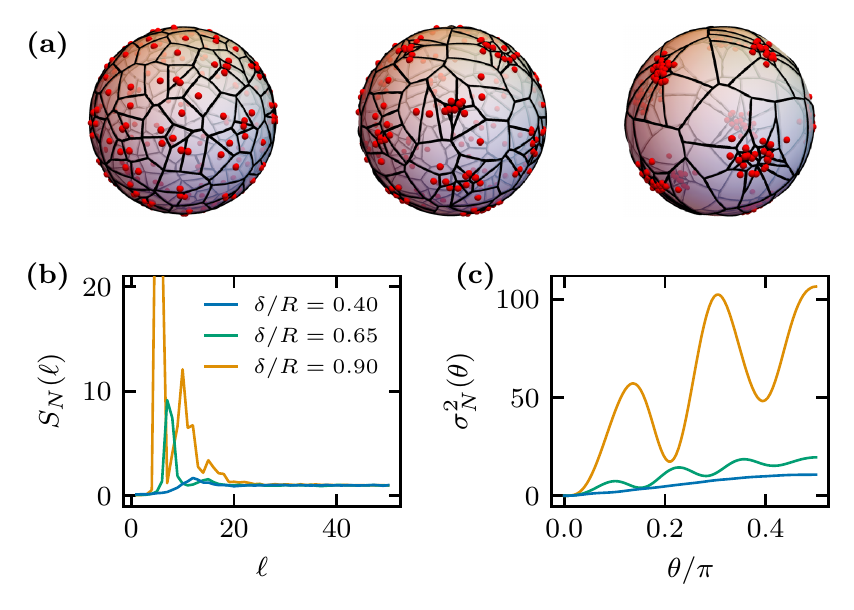}
\caption{{\bf (a)} Voronoi tesselations of distributions of $N=200$ GEM-4 particles at $T=1$ and $\delta/R=0.40$, $0.65$, and $0.90$ (from left to right). Shown are also the ensemble averages of the spherical structure factor {\bf (b)} and cap number variance {\bf (c)} for the same values of $N$, $T$, and $\delta/R$ as in panel (a). The legend of panel (b) also applies to panel (c).\label{fig:3}}
\end{figure}

Different types of order in systems of GEM-4 particles can be clearly seen both in their spherical structure factor and in their cap number variance (Fig.~\ref{fig:3}). When the system is in the homogeneous fluid phase, $S_N(\ell)$ exhibits only a shallow first peak while $\sigma_N^2(\theta)$ shows no modulation related to a shell-like structure, typical for ordered systems. However, when the system is in the cluster crystal phase, $S_N(\ell)$ exhibits several pronounced peaks and $\sigma_N^2(\ell)$ now shows the characteristic modulations related to structural order. Notable is the overall scale of both measures compared to their form: the modulations in $\sigma_N^2(\theta)$ (Fig.~\ref{fig:3}c) is characterized by the number of clusters $N^\ast$ and not the total number of particles $N$, as was the case in the thermal Thomson distributions. At the same time, the large magnitude of the peaks in the spherical structure factor (Fig.~\ref{fig:3}b) when compared to those observed for thermal Thomson distributions (Fig.~\ref{fig:1}b) is due to the fact that each of the $N^\ast$ clusters is composed of $N/N^\ast$ particles on average.

\subsection{Hyperuniformity parameters}

\begin{figure}[!b]
\centering
\includegraphics[width=\columnwidth]{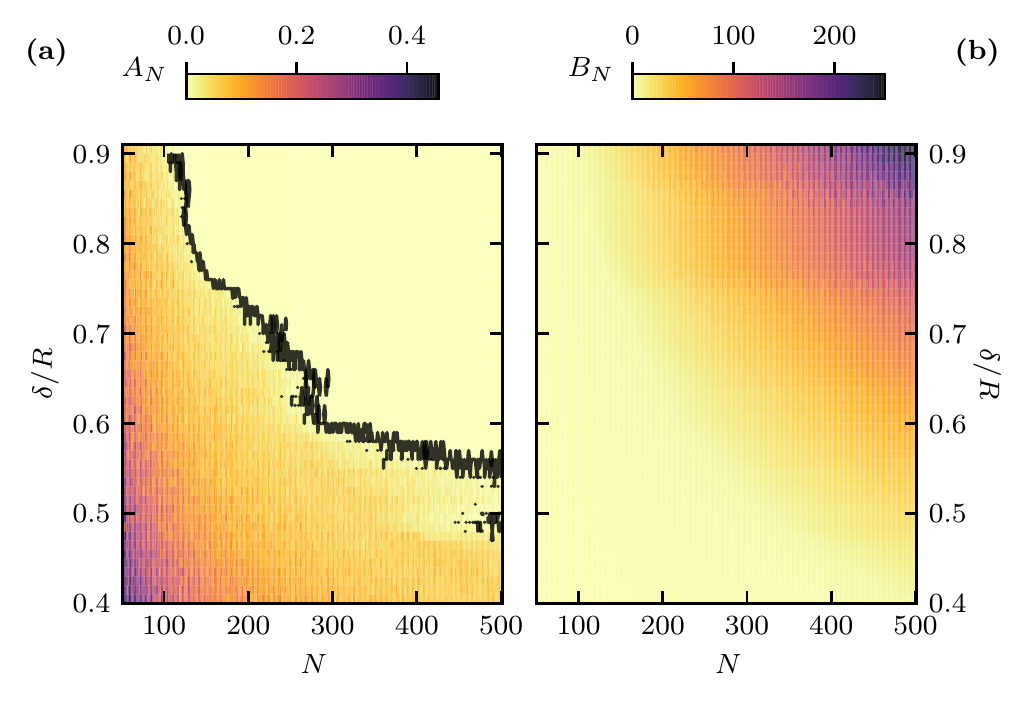}
\caption{Hyperuniformity parameters $A_N$ {\bf (a)} and $B_N$ {\bf (b)} in the $N$-$\delta/R$ plane for distributions of GEM-4 particles at reduced temperature $T=1$. The parameters were obtained as fits of Eq.~\eqref{eq:cnv} to ensemble-averaged cap number variance. Black contour line in panel (a) shows where the parameter $A_N$ vanishes, $A_N\lesssim10^{-10}$.\label{fig:4}}
\end{figure}

While both $S_N(\ell)$ and $\sigma_N^2(\theta)$ show the transition of a system of GEM-4 particles from a homogeneous fluid to a cluster crystal phase, this transition is difficult to capture using standard order parameters due to the disordered nature of particles within each cluster. However, the difference between the two phases is immediately apparent if we take a look at the hyperuniformity parameters $A_N$ and $B_N$, again obtained by fitting Eq.~\eqref{eq:cnv} to the ensemble-averaged $\sigma_N^2(\theta)$. Figure~\ref{fig:4} shows that the value of parameter $A_N$ clearly separates the two phases in the $N$-$\delta/R$ plane. In the homogeneous fluid phase, $A_N$ is always larger than zero. It also never reaches the value of $A_N=1$, indicating that the system is never completely random, which is expected due to the strong interactions between the particles. When the system transitions to the cluster crystal phase, $A_N$ suddenly vanishes, $A_N\lesssim10^{-10}$ (shown by the black region in Fig.~\ref{fig:4}a). This is in stark contrast to the order transition in the thermal Thomson system, where $A_N$ slowly decreased to zero as the temperature was lowered.

When $A_N$ vanishes, $B_N$ again starts to increase. Unlike what we observed in the thermal Thomson problem, or what was previously observed in scale-free distributions~\cite{ALB2019}, $B_N$ can take on extremely large values. The reason is that in the ordered state with $N^\ast$ clusters, the distribution is closer to a hyperuniform distribution of $N^\ast$ particles with larger particle weights, while the expression in Eq.~\eqref{eq:cnv} is normalized only with $N$, as $N^\ast$ is not known in advance. Rescaling the structure factor shows that $B_N$ should scale as a power of the number of particles per cluster, $(N/N^\ast)^{3/2}$. Indeed, this scaling helps explain the observed pattern for the number of clusters in the $N$-$T$ plane~\cite{Franzini2018}, where the number of clusters, and thus the average number of particles per cluster, changes with $\delta/R$ but not with $N$. In the ordered state, the scaling of the parameter $B_N$ shows the same pattern; however, it also includes an unknown prefactor, which we are currently unable to predict theoretically. Nonetheless, the parameter $B_N$ clearly shows the potential to be used for assessing finer aspects of order, such as clustering.

\begin{figure}[!t]
\centering
\includegraphics[width=\columnwidth]{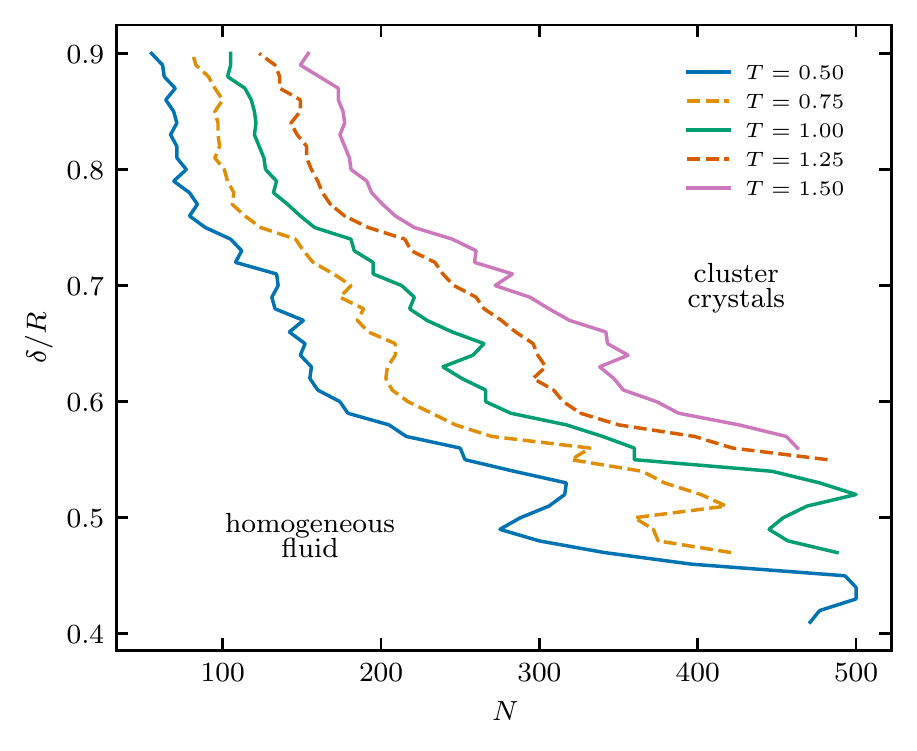}
\caption{Curves of vanishing $A_N$ in the $N$-$\delta/R$ plane for five different temperatures of the system. The curves are defined as points where $A_N\leq10^{-10}$ and mark the transition from a homogeneous fluid in the left part of the phase diagram to a cluster crystal phase in the right part of the diagram.\label{fig:5}}
\end{figure}

The observation that $A_N$ vanishes suddenly with the appearance of cluster crystal phase can be exploited to separate the $N$-$\delta/R$ plane into two regions corresponding to homogeneous fluid and cluster crystal phases. This is shown in Fig.~\ref{fig:5} for five different temperatures of the system. By observing when $A_N\leq10^{-10}$, it is easy to see that the cluster crystals span a larger part of the phase diagram at lower $T$. Moreover, increasing the temperature appears to shift the phase curve towards larger $N$ while maintaining its position in the $\delta/R$ direction. While these observations have been made previously by Franzini et al.~\cite{Franzini2018} in their original study, they did not use an order parameter to delineate the regions of the phase space. Our results demonstrate that $A_N$ and $B_N$ can be used as global order parameters to construct the phase diagram of the system, something which cannot be done using standard order parameters on the sphere.

We also note that there is a larger uncertainty in determining the phase line at low $\delta/R$: the likely reason is that in this regime, a very high number of clusters is formed ($N^\ast\gtrsim50$; cf.\ Ref.~\onlinecite{Franzini2018}), and the transition from a homogeneous fluid to cluster crystal phase becomes blurred. In this part of the phase space our predictions also differ from the observations of Franzini et al.~\cite{Franzini2018}; specifically, we predict that low $\delta/R$ lead to the onset of cluster crystal phase at much higher $N$ than originally thought.

\section{Discussion}

Hyperuniformity has only recently been generalized to non-Euclidean geometries, and the known notions from Euclidean space have been shown to extend to spherical geometry as well. Nonetheless, there are several notable differences between the two, related to the restrictions that topology and geometry of the sphere dictate. At the moment, there seem to be no distinct hyperuniformity classes on the sphere---unlike in the Euclidean case--as the hyperuniformity parameters $A_N$ and $B_N$ derived from cap number variance can change in a continuous manner. However, just as it is already known for scale-free systems of particles on the sphere, we have shown here that these two parameters can be used to consistently detect and study order transitions in systems of particles involving one or more internal length scales.

By studying two such systems---thermal Thomson problem and particles interacting via GEM-4 potential---we have shown that the parameter $A_N$ is a good measure of disorder in the system. While $A_N$ is finite, $A_N\leqslant1$, the system is in either completely disordered ($A_N\approx1$) or fluid-like ($A_N<1$) state. Importantly, when $A_N$ vanishes, $A_N\to0$, the system undergoes an order transition. Once this happens, the parameter $B_N$ becomes relevant and in a crystal-like state takes on a constant value. As $A_N$ gradually vanishes, $B_N$ might also cross a barrier, increasing at first before assuming this value. Both of these observations and the scale of $B_N$ are likely related to the details of the systems, particularly regarding the interactions involved, although further theoretical insights into this are currently still lacking. Ideally, these would connect the exact nature of the interaction potential to an improved approximation of cap number variance, currently given by Eq.~\eqref{eq:cnv}.

When the positions of the particles are known, the fit of Eq.~\eqref{eq:cnv} is easy to carry out not only in simulations but also in experimental realizations of spherical assemblies. However, based on our analysis, we also see that it might be possible to derive some proxies for disorder and the parameter $A_N$ which might be quicker to determine. Such candidates are the dipole moment of the spherical structure factor, $S_N(\ell=1)$, or the hemispherical cap number variance, $\sigma_N^2(\theta=\pi/2$), as both are the first to show drastic changes when a system undergoes an order transition. These are nonetheless likely to fail in systems such as the one involving GEM-4 potential, where only a full analysis of the hyperuniformity parameters $A_N$ and $B_N$ correctly detects the phase transitions of the system.

The results of our study imply that the concept of hyperuniformity on a sphere can be used to study order and order transitions in any system of particles confined to its surface. Future work should aim to incorporate this not only in various simulated systems, but also in experimental situations of, for instance, colloids confined to the surface of liquid droplets. Furthermore, a better theoretical understanding of how the nature of the interactions in a system governs the hyperuniformity parameters $A_N$ and $B_N$ would allow to not only discern between ordered and disordered systems in general but also between the different degrees of order present in them.

\begin{acknowledgments}
This work was funded by Slovenian Research Agency ARRS (Research Core Funding No.\ P1-0055 (A.B.) and No.\ P1-0099 (S.\v C.) and research Grant No.\ J1-9149), and is associated with the COST Action EUTOPIA (Grant No.\ CA17139). S.F.\ acknowledges ``Laboratorio di Calcolo e Multimedia (LCM)'' of the University of Milan for providing machine time on their cluster.
\end{acknowledgments}

\section*{Data availability}

Data that support the findings of this study are available from the corresponding author upon reasonable request.

\bibliography{references}% Produces the bibliography via BibTeX.

\end{document}